\documentclass[aps,prc,notitlepage,floatfix,nofootinbib,amsmath,amssymb,amsfonts,twocolumn]{revtex4-2}

\usepackage{graphicx}
\usepackage{hyperref}
\usepackage{multirow}
\newcommand{\DCSB}{$\mathcal{D}_{{\rm CSB}}$\,}
\newcommand{\DCS}{$\mathcal{D}_{{\rm CS}}$\,}

\begin{document}

\title{Performing Bayesian analyses with \texttt{AZURE2} using \texttt{BRICK}:
an application to the \texorpdfstring{{}${}^7$}{7}Be system}

\author{Daniel Odell}
	\affiliation{Institute of Nuclear and Particle Physics and Department of Physics and Astronomy, Ohio University, Athens,
		Ohio 45701, USA}

\author{Carl~R. Brune}
	\affiliation{Institute of Nuclear and Particle Physics and Department of Physics and Astronomy, Ohio University, Athens,
		Ohio 45701, USA}

\author{Daniel~R. Phillips}
	\affiliation{Institute of Nuclear and Particle Physics and Department of Physics and Astronomy, Ohio University, Athens,
		Ohio 45701, USA}
		
\author{Richard James deBoer}
    \affiliation{Department of Physics and the Joint Institute for Nuclear Astrophysics, University of Notre Dame, Notre Dame, Indiana 46556 USA}
    
\author{Som Nath Paneru}
    \affiliation{Facility for Rare Isotope Beams,
    Michigan State University, East Lansing, Michigan, USA (\textit{present})}
    \affiliation{Institute of Nuclear and Particle Physics and Department of Physics and Astronomy, Ohio University, Athens,
		Ohio 45701, USA}

\date{\today}

\begin{abstract}
Phenomenological $R$-matrix has been a standard framework for the evaluation of resolved resonance cross section data in nuclear physics for many years. It is a powerful method for comparing different types of experimental nuclear data and combining the results of many different experimental measurements in order to gain a better estimation of the true underlying cross sections. Yet a practical challenge has always been the estimation of the uncertainty on both the cross sections at the energies of interest and the fit parameters, which can take the form of standard level parameters. Frequentist ($\chi^2$-based) estimation has been the norm.  In this work, a Markov Chain Monte Carlo sampler, \texttt{emcee}, has been implemented for the $R$-matrix code \texttt{AZURE2}, creating the Bayesian $R$-matrix Inference Code Kit (\texttt{BRICK}). Bayesian uncertainty estimation has then been carried out for a simultaneous $R$-matrix fit of the $^3$He$(\alpha,\gamma)^7$Be and $^3$He$(\alpha,\alpha)^3$He reactions in order to gain further insight into the fitting of capture and scattering data. Both data sets constrain the values of the bound state $\alpha$-particle asymptotic normalization coefficients in $^7$Be. The analysis highlights the need for low-energy scattering data with well-documented uncertainty information and shows how misleading results can be obtained in its absence. 
\end{abstract}

\maketitle


\section{Introduction}

Phenomenological $R$-matrix has been the standard analysis tool for cross section data that exhibit overlapping yet resolved resonances for many years~\cite{Lan58}. It is used extensively to evaluate data for applications (e.g. the ENDF/B-VIII.0 evaluation~\cite{Bro18}), to perform extrapolations to low, unobserved energies in nuclear astrophysics (e.g. \citet{Azu10, Des05}), and to extract level parameters for nuclear structure. In all cases, it provides a reaction framework in which experimental information of various different types can be combined to improve estimates of the true cross sections. One challenging aspect of this type of analysis has been reliable uncertainty propagation.

Traditionally, data have been fitted using $\chi^2$ minimization, with uncertainties being estimated using one of two methods. The first is using partial derivatives and the assumption that the quantity of interest is related linearly with the parameters of the model. The second is the assignment of confidence intervals  based on some $\Delta\chi^2$ value. The assumption of linearity is often a poor one and the second method can become tedious or impossible to implement for a complicated model. Additional limitations are that one must assume Gaussian uncertainties on the input data and there is almost no ability to include prior information about the parameters. It is known that $\chi^2$ methods may lead to biased results and/or underestimated uncertainties in data evaluations~\cite{Smith_2007}. The reason for these issues is understood to be incomplete documentation or modeling of systematic uncertainties. While systematic uncertainties are a difficult subject in any approach, they are much easier to model and implement using the Bayesian methods described below. Finally, we would like to point out that a mixed approach is possible, where $\chi^2$ minimization is combined with a Monte Carlo simulation of some uncertainties. This method was used by \textcite{deBoer:2014hha} in a previous analysis of $^3$He$(\alpha,\gamma)^7$Be and $^3$He$(\alpha,\alpha)^3$He.


Bayesian methods are increasingly becoming the standard for performing Uncertainty Quantification in physical sciences and engineering in general, and theoretical nuclear physics in particular~\cite{Schindler:2008fh,Furnstahl:2014xsa,Furnstahl:2015rha,Zhang:2015ajn,Melendez:2017phj,Wesolowski:2018lzj,Neufcourt:2019sle,Neufcourt:2019qvd,King:2019sax,Melendez:2019izc,Filin:2019eoe,Drischler:2020hwi,Drischler:2020yad,Premarathna:2019tup,Zhang:2019odg,Filin:2020tcs,Schunck:2020lvn,Neufcourt:2020nme,JETSCAPE:2020mzn,Catacora-Rios:2020xgx,Reinert:2020mcu,Phillips:2020dmw,Wesolowski:2021cni,Schnabel:2021tzq,Xu:2021aij,JETSCAPE:2021ehl,Hamaker:2021nzy}. In contrast to a traditional $\chi^2$-minimization they offer the opportunity to examine the entire probability distribution for parameters of interest, rather than focusing on the values that maximize the likelihood. Perhaps equally important, in a Bayesian approach it is straightforward---mandatory even---to declare and include prior information on the parameters of interest. Bayesian methods, combined with the possibility to use Markov Chain Monte Carlo sampling to explore a high-dimensional parameter space, allow one to introduce additional parameters without fear of computational instabilities caused by shallow $\chi^2$ minima. The use of MCMC sampling also makes uncertainty propagation straightforward, as we will demonstrate here. And a Bayesian  framework is---to our knowledge---the only option if one wishes to incorporate a rigorous formulation of theory uncertainties into the statistical analysis. In this work, Bayesian uncertainty quantification is implemented by pairing the $R$-matrix code \texttt{AZURE2}~\cite{Azu10, Ube15} with the MCMC Python package \texttt{emcee}~\cite{For13}. The pairing is facilitated by a Python interface \texttt{BRICK} (Bayesian R-matrix Inference Code Kit), enabling Bayesian inference in the context of $R$-matrix analyses.

\begin{figure*}
    \centering
    \includegraphics{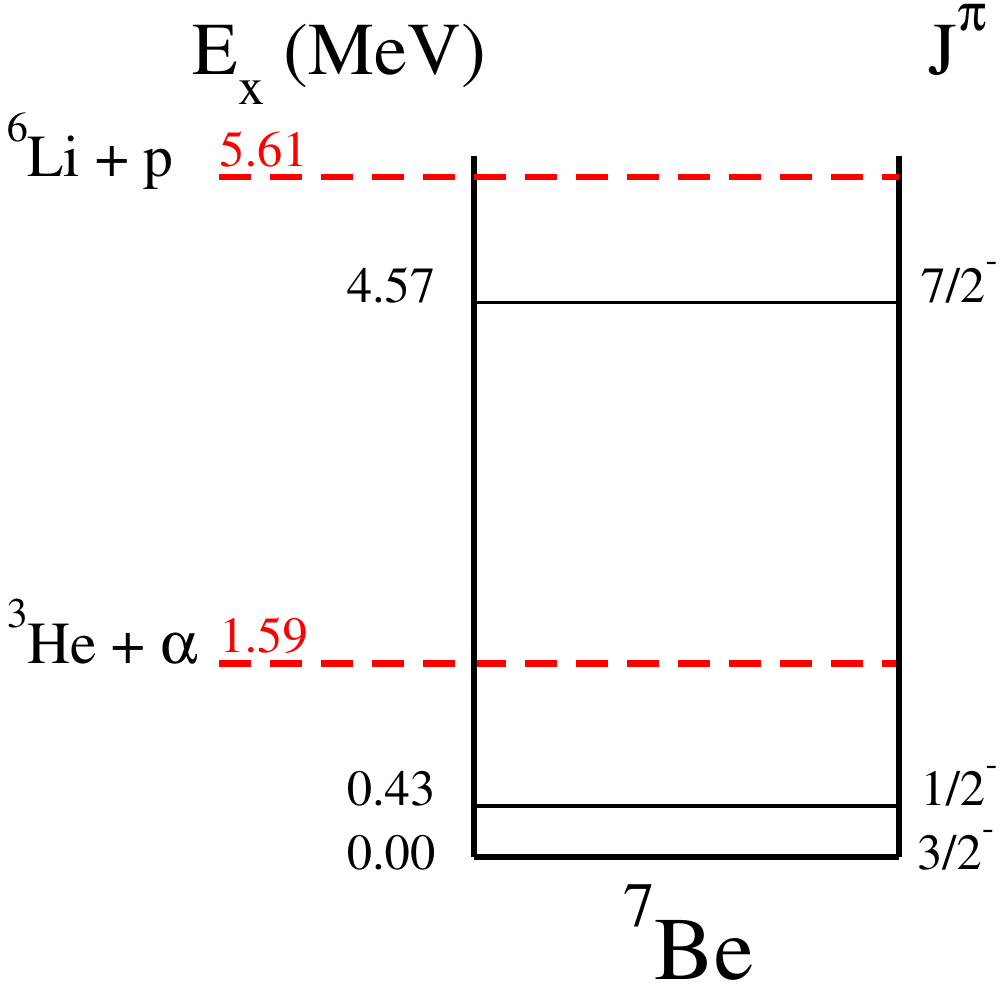}
    \caption{Level diagram of $^7$Be up to the proton separation energy.}
    \label{fig:level_diagram}
\end{figure*}

To benchmark this code, it has been applied to the analysis of the $^3$He$(\alpha,\gamma)^7$Be and $^3$He$(\alpha,\alpha)^3$He reactions. The $^3$He$(\alpha,\gamma)^7$Be reaction is a key reaction in modeling the neutrino flux coming from our sun~\cite{RevModPhys.60.297}. It also plays a role in Big Bang Nucleosynthesis (BBN). The reaction cross section is dominated by the direct capture process, but also has significant contributions from broad resonances (see Fig.~\ref{fig:level_diagram}). In recent years, high-precision measurements of this reaction have been performed, using direct $\gamma$-ray detection~\cite{Kon13, Bro07, Cos08}, the activation method~\cite{Sin04, Bro07, Car12, Cos08, Bor13}, and a recoil separator~\cite{Dil09}. Additional higher energy measurements have also been made recently by \citet{Szu19}, but are outside the energy range of the present analysis. Using these high precision measurements, several analyses have been made to combine these data sets and extrapolate the cross section to low energies using pure external capture~\cite{Ade11}, $R$-matrix~\cite{deBoer:2014hha}, effective field theory~\cite{Zhang:2019odg, Rup20}, a modified potential model~\cite{Tur21}, and {\it ab initio} calculations~\cite{Nollett:2001ub,Nef11,Dohet-Eraly:2015ooa,Vorabbi:2019imi}. These several recent analyses make this reaction an ideal case for benchmarking since they employ both more traditional and Bayesian uncertainty estimation methods.

As the energies pertinent to solar fusion and BBN the $^3$He$(\alpha,\gamma)^7$Be cross section has a large contribution from external capture, $^3$He$(\alpha,\alpha)^3$He data, through its constraints on the scattering phase shifts, should also provide an additional source of constraint on the low-energy extrapolation. This type of combined analysis has been reported in \citet{deBoer:2014hha}, but there it was found that the available scattering data of \citet{Bar64} was inconsistent with the capture data, perhaps because of incomplete uncertainty documentation in the former. With this in mind, new measurements of the $^3$He$(\alpha,\alpha)^3$He cross section were recently reported by \citet{the_Paneru}. 

In this work, a Bayesian uncertainty analysis is performed on an $R$-matrix fit to the low energy $^3$He$(\alpha,\gamma)^7$Be~\cite{Kon13, Cos08, Bro07, Sin04, Car12, Bor13, Dil09} and $^3$He$(\alpha,\alpha)^3$He~\cite{Bar64, the_Paneru} data, including the new SONIK data. The sensitivity of the fit to the scattering data is the main focus, examining the differences resulting from the two different scattering data sets considered. The mapping of the posterior distributions of the fit parameters, cross sections, phase shifts, and scattering lengths gives new insights into the dependence of these quantities to the input scattering data.


    
    
    

\section{What is \texttt{BRICK}?}

\texttt{BRICK} is a python package that acts as an interface between the \texttt{AZURE2}~\cite{Azu10, Ube15} $R$-matrix code and an MCMC sampler.
It is not a replacement for \texttt{AZURE2} nor is it intended to be.
The primary functionality that it provides is a user-friendly way to sample parameters that have already been set up with the \texttt{AZURE2} GUI to be varied.

\subsection{\texttt{AZURE2}}

\texttt{AZURE2} is a multilevel, multichannel, $R$-matrix code (open source) that was developed under the Joint Institute for Nuclear Astrophysics (JINA)~\cite{Azu10, Ube15}. While the code was created primarily to handle the added complexity of charged-particle induced capture reactions~\cite{Deb17}, also has capability for a wide range of other types of reaction calculations. The code is primarily designed to be used by way of a graphical user interface (GUI), but can also be executed in a command line mode for batch processes~\cite{Ube15}. The code stores all of its setup information in a simple text input file. While this file is usually edited by way of the GUI, it can also be modified directly. This may be desirable for batch type calculations, as are being used here.

\texttt{AZURE2} primarily uses the alternative $R$-matrix parameterization of \citet{Bru02}. It has two main advantages. The first is that it eliminates the need for the boundary conditions present in the classical formalism of \citet{Lan58}. The second is that the remaining fit parameters become the observed level parameters. The remaining model parameters are the channel radii.

A key advantage in using the parameterization of \citet{Bru02} for the fitting of low energy capture reactions is that level parameters for bound or near threshold resonances can be more directly included in the $R$-matrix analysis~\cite{Muk99,Muk01}. The use of the Bayesian uncertainty estimation further facilitates the inclusion of uncertainty information for these parameters. This provides an improved method for communicating the level structure information gained from transfer reaction studies into an $R$-matrix analysis in a statistically rigorous way.




\subsection{Implementation}

\begin{figure*}
    \centering
    \includegraphics{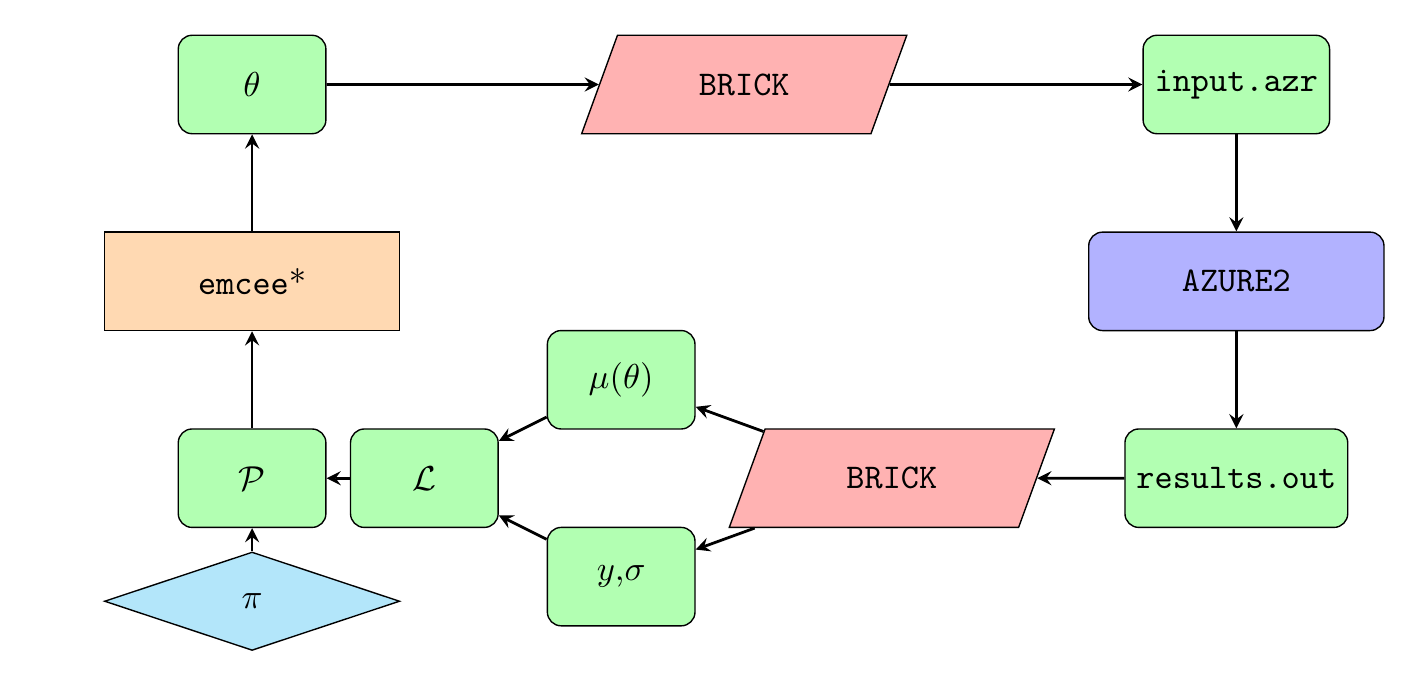}
    \caption{Representation of the different roles of \texttt{emcee}, \texttt{BRICK}, and \texttt{AZURE2} in the Bayesian analysis presented below. The asterisk in the \texttt{emcee} rectangle indicates the starting point of the process.}
    \label{fig:implementation}
\end{figure*}

\paragraph{Overview}
The role of \texttt{BRICK} in our $R$-matrix calculations is to act as a mediator.
It maps proposed parameters --- both $R$-matrix parameters and normalization factors --- from an MCMC sampler to \texttt{AZURE2} and $R$-matrix predictions from \texttt{AZURE2} back to the sampler.
First, it accepts proposed points in parameter space, $\theta$, from the sampler --- in this analysis we use \texttt{emcee}~\cite{For13} --- and packages them into a format that \texttt{AZURE2} can read.
Then it reads the output from \texttt{AZURE2} and presents it as a list.
Each item of the list contains the predictions, $\mu(\theta)$, and data, $y$ and $\sigma$, corresponding to a specific output channel configuration.
The likelihood can then be calculated according to the user's choice, see
 \eqref{eq:likelihood} for one possibility.
Accompanied by user-defined priors, one can readily construct a Bayesian posterior.
This posterior value, $\ln\mathcal{P}$, is passed back to \texttt{emcee}.
Finally, based on the $\ln\mathcal{P}$ value, the MCMC algorithm proposes a new $\theta$ and the process repeats.
A diagram is provided in Fig.~\ref{fig:implementation} to illustrate the qualitative functionality of the different software packages.
The process described above starts at the orange rectangle labeled ``\texttt{emcee}''.


\paragraph{Details}

\texttt{BRICK} is built such that different samplers can be used.
The analysis presented in this paper uses \texttt{emcee}, so the details provided in this section will be somewhat specific to it.

When initializing an instance of an \texttt{EnsembleSampler}, the most relevant argument is \texttt{log\_prob\_fn}, the function that returns the logarithm of the probability.
One of the advantages of \texttt{emcee} is that it allows the practitioner to perform arbitrary calculations inside that probability function.
That function must meet only two requirements: (1) take an array of floating point numbers that represents the vector in parameter space and (2) return a floating point number that represents the logarithm of the probability associated with that array.
In between those two steps, one is free to perform whatever calculations one needs.
This can be seen on the left-hand side of Fig.~\ref{fig:implementation}.
The parameter-space vector, $\theta$, is output from \texttt{emcee}.
The logarithm of the probability at that point, $\ln\mathcal{P}$, is subsequently input to \texttt{emcee}.
In this sense, \texttt{emcee} is well-suited to the implementation of ``black-box'' physics models where one has limited access to the source code.

The primary tasks that \texttt{BRICK} accomplishes are (1) translating $\theta$ into a format that \texttt{AZURE2} can read and (2) reading the output from \texttt{AZURE2} such that a $\ln\mathcal{P}$ value can be easily calculated.
The means of accomplishing these tasks relies on the command-line interface (CLI) to \texttt{AZURE2}, which is accessible when installed on Linux machines.
The CLI options available to \texttt{AZURE2} are well documented in the manual~\cite{Ube15}.
The most critical argument is the input file, typically accompanied by the file extension \texttt{.azr}.
This input file contains all of the necessary information to perform an $R$-matrix calculation with a given set of parameters.
It is generated when the $R$-matrix and data models are built with the commonly used GUI, which \texttt{AZURE2} provides.
\texttt{BRICK} is not built to replace that GUI.
It \textit{accompanies} \texttt{AZURE2} by allowing the user to bring their \texttt{AZURE2}-prepared $R$-matrix model over and sample what was previously optimized.
Accordingly, the default behavior of \texttt{BRICK} is to respect the choices made by the user in the \texttt{AZURE2} GUI.
If a parameter is fixed in \texttt{AZURE2}, it is fixed in \texttt{BRICK}.
If it is varied in \texttt{AZURE2}, it is sampled in \texttt{BRICK}.

\texttt{BRICK} accesses the \texttt{AZURE2} CLI through the Python module \texttt{subprocess}.
But prior to that, \texttt{BRICK} must map the values in $\theta$ to the proper locations in the input file.
This is accomplished by reading the \texttt{<levels>} and \texttt{<sectionsData>} sections of the input file.
\texttt{BRICK} reads the appropriate parameters and flags looking for varied parameters.
As they are found, their locations are stored.
When a new $\theta$ is proposed, \texttt{BRICK} creates a new input file and maps the values in $\theta$ to the varied parameter locations.
Then \texttt{AZURE2} is called with the newly generated input file.
The output from \texttt{AZURE2} is written to a sequence of files in the \texttt{output} directory by default.
Those files are read and the predictions, $\mu$, and experimental data, $y$, are extracted.
A likelihood is then constructed.
Under the assumption that the uncertainties associated with $y$ are uncorrelated and normally distributed, this is a multivariate Gaussian distribution.
Accompanied by a list of prior distributions corresponding to the preexisting knowledge of the sampled parameters, a posterior is finally constructed and passed back to \texttt{emcee}.

Initially, this process was built in a single-threaded manner.
As \texttt{emcee} is a ensemble sampler, efficient exploration of the posterior relies heavily on many, simultaneous walkers.
In order to scale this beyond the most basic calculations, we modified our implementation to allow each walker to write its own input file and read from its own output directory.
Inside the log-probability function, there is no access to any kind of walker identifier, so each walker generates a file-space that is uniquely identified by an eight-character random string.
This allows each walker to work independently, so on systems where many cores are available, each walker can have a dedicated core.
Or at least the time spent waiting for CPU time is minimized.
This also allows for an increased number of walkers, which is a common tactic used to decrease autocorrelation time.

\section{Application to \texorpdfstring{$^3$H\MakeLowercase{e}$(\alpha,\alpha)^3$H\MakeLowercase{e}}{3He(a,a)} and \texorpdfstring{$^3$H\MakeLowercase{e}$(\alpha,\gamma)^7$B\MakeLowercase{e}}{3He(a,g)}}

\subsection{The \texorpdfstring{$R$}{R}-matrix model}

The starting point for the $R$-matrix model used here is that of \citet{deBoer:2014hha}. In that work, ten levels were used with three particle pairs ($^3$He+$\alpha$, $^7$Be+$\gamma_0$, and $^7$Be+$\gamma_1$) for a total of 16 $R$-matrix fit parameters. Initial MCMC calculations showed that a 7/2$^-$ background level was not statistically significant, and was thus dropped from the calculation. This already demonstrates one of the powerful feature of this type of MCMC analysis, it provides a clear identification of redundant fit parameters. The $R$-matrix model used here thus consisted of nine levels, three particle pairs, and 16 $R$-matrix fit parameters as summarized in Table~\ref{tab:sampled_rpars}. 

\begin{table*}[]
    \centering
    \begin{ruledtabular}
    \caption{Sampled parameters in the $R$-matrix model. Numbers indicate that the level energies were fixed. A variable indicates that the parameter was sampled. The subscripts $\alpha$ and $\gamma$ indicate the exit particle pair --- scattering and capture, respectively. Capture particle pairs are distinguished by ground (0) and excited (1) ${}^7\rm{Be}$ states. \label{tab:sampled_rpars}}
    \begin{tabular}{c|c|c|c}
        $J^\pi$ & $E_\lambda$ ($E_x$, MeV) & Widths and ANCs & Prior Distributions \\
        \hline
         $1/2^-$ & 0.4291 & $C_1$ & $U(1, 5{\rm MeV})$ \\
         $1/2^-$ & 21.6 & $\Gamma_\alpha$ & $U(-200, 200{\rm MeV})$ \\
         \hline
         \multirow{3}{*}{$1/2^+$} & \multirow{3}{*}{14} & 
         $\Gamma_\alpha$ & $U(0, 100{\rm MeV})$ \\ 
         & & $\Gamma_{\gamma,0}$ & $U(0,10{\rm MeV})$\\
         & & $\Gamma_{\gamma,1}$ & $U(-10,10{\rm keV})$\\
         \hline
         $3/2^-$ & 0 & $C_0$ & $U(1,5{\rm MeV})$ \\
         \hline
         $3/2^-$ & 21.6 & $\Gamma_\alpha$ & $U(-100,100{\rm MeV})$ \\
         \hline
         \multirow{3}{*}{$3/2^+$} & \multirow{3}{*}{12} & $\Gamma_\alpha$ & $U(0,100{\rm MeV})$ \\
         & & $\Gamma_{\gamma,0}$ & $U(-10,10{\rm keV})$ \\
         & & $\Gamma_{\gamma,1}$ & $U(-3,3{\rm keV})$ \\
         \hline
         $5/2^-$ & 7 & $\Gamma_\alpha$ & $U(0,100{\rm MeV})$ \\
         \hline
         \multirow{2}{*}{$5/2^+$} & \multirow{2}{*}{12} & $\Gamma_\alpha$ & $U(0,100{\rm MeV})$ \\
         & & $\Gamma_{\gamma,0}$ & $U(-100,100{\rm MeV})$ \\
         \hline
         \multirow{2}{*}{$7/2^-$} & \multirow{2}{*}{$E^{(7/2^-)}$ [$U(1,10{\rm MeV})$]} & $\Gamma_\alpha$ & $U(0,10{\rm MeV})$ \\
         & & $\Gamma_{\gamma,0}$ & $U(0,1{\rm keV})$
    \end{tabular}    
\end{ruledtabular}
\end{table*}

\subsection{Priors on \texorpdfstring{$R$}{R}-matrix parameters}\label{sec:priors}

Because this is a Bayesian analysis, we must choose priors for all $R$-matrix parameters. We have chosen to employ uninformative uniform priors. However, the signs of the reduced width amplitudes (that is the interference solution), which are implemented in \texttt{AZURE2} by the signs of the partial widths, were determined by the initial best $\chi^2$ fit using \texttt{AZURE2}. In this case, a unique interference solution was found. This may not always be the case: sometimes other interference solutions may be possible. The \texttt{emcee} sampler may then not be able to easily find these other interference solutions in the parameter space. It seems is likely that in cases where different interference solutions are possible, each one will require a separate \texttt{emcee} analysis.

One common circumstance where a Bayesian analysis will improve on previous uncertainty estimates is in the ability to give priors for bound state level parameters determined from transfer studies. Unfortunately, in the case of the $^7$Be system, there is limited information available for the bound state $\alpha$-particle ANCs. A recent first measurement has been reported by \citet{Kis20}, but the ANCs are rather discrepant from those found from this and past $R$-matrix analyses of capture data. This inconsistency has not been investigated here, but needs to be addressed in future work. If reliable bound-state ANC determinations become available, that are independent of the capture and scattering data, it provides a path to further decrease the uncertainty in the low energy $S$-factor extrapolation. One could also adopt priors on the ANCs from {\it ab initio} calculations, although we have not done so.

It is also tempting to implement more constraining priors into the $R$-matrix analysis from a compilation like the National Nuclear Data Center or the TUNL Nuclear Data Project~~\cite{TILLEY20023}. However, great care must be taken to carefully understand the source of the values and uncertainties when weighted averages are used to determine adopted values for level parameters in these compilations. In particular, past analysis of the data being fit in the $R$-matrix analysis may be a contributor to the evaluation values. Thus blindly using evaluation level parameters and uncertainties can lead to double counting and an erroneous decrease of uncertainties. It is for this reason that uniform priors on parameters are adopted in the present analysis. The posterior shapes then clearly stem solely from the data sets considered in the $R$-matrix analysis.


The priors for the $R$-matrix parameters used in this work are listed in Table~\ref{tab:sampled_rpars}. 
In most cases, level energies are fixed.
For details of the choices made in formation of the $R$-matrix model, see~\textcite{PaneruThesis}.
The distribution formed by the product of these $R$-matrix priors and priors on the parameters introduced in the next section is the overall prior $\pi$ shown in Fig.~\ref{fig:implementation}. 

\subsection{Modeling systematic errors in the data}

\label{sec:systematic_errors}

\paragraph{Common-mode errors}

\texttt{AZURE2} provides a method for the inclusion of a common-mode error for each data set using a modified $\chi^2$ function
\begin{equation}
        \label{eq:chi_squared}
        \chi^2=\sum_{\alpha=1}^{N_{\rm sets}}\left(\sum_{j=1}^{N_\alpha} \frac{(f(x_{\alpha,j})-c_{\alpha}n_{\alpha}y_{\alpha,j})^{2}}{(c_{\alpha}n_{\alpha}\sigma_{\alpha,j})^2}+\frac{((c_{\alpha}-n_{\alpha})/n_{\alpha})^2}{\delta_{c_{exp},\alpha}^{2}}\right),
    \end{equation}
where $c_{\alpha}$ is the normalization fit parameter, $n_{\alpha}$ is the starting normalization which is set to 1 in the present analysis, $f(x_{\alpha,j})$ is the differential scattering cross section form the $R$-matrix, $y_{\alpha,j}$ is the data point value, $\sigma_{\alpha,j}$ is the combined statistical and point-to-point uncertainty of a data point, and $\delta_{c_{exp},\alpha}$ is the {\it fractional} common-mode uncertainty of the data set. The additional term in the $\chi^2$ function is derived by making the approximation that the common-mode systematic uncertainty has a Gaussian probability distribution~\cite{DAGOSTINI1994306}. The accuracy of this approximation is often unclear.

Common-mode errors are implemented in the present analysis in \texttt{BRICK}, outside of \texttt{AZURE2}, i.e., the common-mode errors are applied to the \texttt{AZURE2} output. In \texttt{BRICK} the $R$-matrix parameter set $\theta_R$ is augmented by a set of normalization factors $f_\alpha$ and energy shifts, $\Delta_{E,\alpha}$. (At present energy shifts are only implemented for scattering data.)
The overall parameter set $\theta$ is then the union of the set $\theta_R$ and $\{f_\alpha,\Delta_{E,\alpha}\}$. The likelihood ${\cal L}$ is formed as a product of standard Gaussian likelihoods for each data point, but with normalization factors applied to the \texttt{AZURE2} predictions $\mu$:
\begin{equation}
    {\cal L} \propto \prod_{\alpha=1}^{N_{\rm sets}} \prod_{j=1}^{N_\alpha}
    \exp\left(-\frac{(y_{j \alpha} - f_\alpha\mu(x_{j \alpha} + \Delta_{E,\alpha};\theta_R))^2}{2 \sigma_{j \alpha}^2}\right),
    \label{eq:likelihood}
\end{equation}
where we have omitted overall factors that do not affect the parameter estimation. Here $x_{j \alpha}$ represents the kinematics of the $j$th data point in data set $\alpha$, and $\sigma_{j \alpha}$ is the combined statistical and point-to-point uncertainty of the corresponding datum, $y_{j \alpha}$. $N_\alpha$ is the number of points in data set $\alpha$, and the product over $\alpha$ runs over all the sets that have independent common-mode errors. 

The priors on the $f_\alpha$'s are specified by the \texttt{BRICK} user. If a Gaussian prior centered at 1 with a width equal to the common-mode error reported in the original experimental publication is employed for the $f_\alpha$'s, then the product of that prior on the normalization factors and the likelihood (\ref{eq:likelihood}) has the same maximum value as the ``extended likelihood" corresponding to \eqref{eq:chi_squared}, that is used to estimate the $f_\alpha$'s in the frequentist framework implemented in \texttt{AZURE2}.

In our analysis of the ${}^3$He($\alpha$,$\alpha$)${}^3$He and ${}^3$He($\alpha$,$\gamma$)${}^7$Be reactions, we adopted such a Gaussian prior, truncated to exclude negative values of the cross section. We used a different $f_\alpha$ for each energy bin in the SONIK data, detailed in Section~\ref{sec:sonik_data}, with the widths of the prior given by the common-mode errors stated in Table~\ref{tab:cmesonik}. The common-mode error associated with the Barnard data, described in Section~\ref{sec:barnard_data}, is taken to be 5\%. The width of the priors for the $f_\alpha$'s to be applied to the capture data, discussed in Section~\ref{sec:capture_data}, are specified by the common-mode errors listed in Table~\ref{tab:cmecapture}.
All normalization-factor priors are of the form
\begin{equation}
    \label{eq:norm_factor_prior}
    T(0,\infty)N(1,\sigma_{f\alpha}^2)~,
\end{equation}
where 
\begin{equation}
    T(a,b) =
    \begin{cases}
    1 & [a, b] \\
    0 & {\rm otherwise}~,
    \end{cases}
\end{equation}
and $N(\mu,\sigma^2)$ represents a Gaussian distribution centered at $\mu$ with a variance of $\sigma^2$.

\begin{table*}[]
    \centering
    \begin{ruledtabular}
    \begin{tabular}{c|c|c}
        Energy (keV/u) & No of Data Points & Common-Mode Errors \\
        \hline
        239 & 17 & 6.4 \\
        291 &  29 & 7.6 \\
        432 & 45 & 9.8 \\
        586 & 46 & 5.7 \\
        711 & 52 & 4.5 \\
        873${}^{(1)}$ & 52 &6.2 \\
        873${}^{(2)}$ & 52 &4.1 \\
        1196 & 52 &  7.7 \\
        1441 & 53 &  6.3 \\
        1820 & 53 & 8.9
    \end{tabular}
    \caption{Common-mode errors associated with the SONIK measurements included in both \DCS\, and \DCSB\, analyses.}
    \label{tab:cmesonik}
    \end{ruledtabular}
\end{table*}

\begin{table*}[]
    \centering
    \begin{ruledtabular}
    \begin{tabular}{c|c|c|c}
        Data Set & Total Capture & Branching Ratio & Common-Mode Errors (\%) \\
        \hline
         Seattle~\cite{Bro18} & 8 pts [0.57, 2.17 MeV] & 8 pts [0.57, 2.17 MeV] & 3 \\
         Weizmann~\cite{Sin04} & 4 pts [0.74, 1.67 MeV] & - & 3.7 \\
         LUNA~\cite{Cos08} & 7 pts [0.16, 0.30 MeV] & 3 pts [0.17, 0.30 MeV] & 3.2 \\
         ERNA~\cite{Dil09} & 47 pts [1.23, 5.49 MeV] & 6 pts [1.93, 4.55 MeV] & 5\\
         Notre Dame~\cite{Kon13} & 17 pts [0.53, 2.55 MeV] & 17 pts [0.53, 2.55 MeV] & 8 \\
         ATOMKI~\cite{Bor13} & 5 pts [2.58, 4.43 MeV] & - & 6
    \end{tabular}
    \caption{Capture data included in both \DCS\, and \DCSB\, analyses: number of data points, energy ranges, and common-mode errors. Energies are given the laboratory frame.}
    \label{tab:cmecapture}
    \end{ruledtabular}
\end{table*}

\paragraph{Energy shifts} 

\texttt{BRICK} also has the capability of estimating (overall) beam-energy shifts in a particular data set. This is implemented as another parameter to be estimated $\Delta_{E,\alpha}$. This parameter affects all the \texttt{AZURE2} evaluations for data set $\alpha$. \texttt{BRICK} implements the energy shift by generating a different input and data files for each value of $\Delta_{E,\alpha}$ under consideration. The flowchart of Fig.~\ref{fig:implementation} is thus not strictly accurate when this feature is included.
Gaussian priors were defined, centered at zero, on possible energy shifts for the SONIK data and the Barnard data. The widths of the priors are based on information in the original papers, as summarized in Sections~\ref{sec:barnard_data} and~\ref{sec:sonik_data}. For the SONIK data, the energy-shift parameter's prior has a standard deviation of 3~keV, based on the energy uncertainty quoted in \citet{the_Paneru}. \citet{Bar64} cites a much larger uncertainty of 20-40~keV, depending upon the energy. The standard deviation of the prior on the $\Delta E$ parameter is taken to be 40~keV for this data set, a much larger value than for the SONIK data. It should be noted that the energy uncertainty for the Barnard data set is not a constant, but it is not possible to improve our modeling of this uncertainty due to the lack of documentation of its origin.

\section{Data sets}

\subsection{\texorpdfstring{\citet{Bar64}}{Bar64} \texorpdfstring{${}^3$}{3}He-\texorpdfstring{$\alpha$}{alpha} elastic scattering}
\label{sec:barnard_data}

Measurements of the elastic scattering products resulting from a ${}^3{\rm He}$ beam incident on a ${}^4{\rm He}$ target were reported in 1964 by \textcite{Bar64}, for $2.4\le E[{}^3{\rm He}, {\rm lab}]\le 5.7$~MeV ($1.4\le E_{\rm c.m.} \le 3.3$~MeV).
The experiment provides excitation functions of differential cross section at eight center-of-mass (c.m.) angles covering $31.55^\circ\le\theta[{}^3{\rm He}, {\rm lab}]\le 91.94^\circ$ ($54.77^\circ \le \theta_{\rm c.m.} \le 140.8^\circ$).
The systematic uncertainty in the measurements is estimated to be 5\%. Detailed point-to-point uncertainties are not given, but are stated to be about 3\%. The measurements are subject to a significant energy uncertainty, estimated to be 20~keV below $E[{}^3{\rm He}, {\rm lab}]=4$~MeV and 40~keV above that energy. It was also noted by the authors that their beam energy was only reproducible to the level of 20~keV.
In total, there are 646 data points collected at 577 unique energies.
The data were obtained from EXFOR in the fall of 2021 and converted into the laboratory frame when necessary.
All eight angles were included. The previous analysis by \citet{deBoer:2014hha} omitted the largest angle.

\subsection{Paneru {\it et al.} \texorpdfstring{${}^3$He-$\alpha$}{} elastic scattering}
\label{sec:sonik_data}
A new measurement of $^3$He+$\alpha$ elastic scattering was performed at TRIUMF using the Scattering of Nuclei in Inverse Kinematics (SONIK)~\cite{Devin,SNPaneru} target and detector system. SONIK was filled with $^4\text{He}$ gas maintained at a typical pressure of 5~Torr bombarded with $^3$He with a beam intensity of about 10$^{12}$ pps. Elastic scattering cross sections were measured at nine different energies from $E_{\text{c.m.}}$~=~0.38--3.13~MeV. SONIK covers an angular range of $30^{\circ}<\theta_{\text{c.m.}}<139^{\circ}$---a markedly larger range than previous measurements. The detectors in SONIK were arranged such that they observed three different points, termed interaction regions, in the gas target along the beam direction. When the beam traversed the gas target it lost energy, so the bombarding energy, and therefore  the  scattering energy, was slightly different in each of the three interaction regions. 

As we will explore further below, the results for the differential scattering cross section from this measurement are consistent with previous determinations but have better precision. The data also extend to markedly lower energies. The uncertainties with this measurement are well quantified and are presented in \citet{the_Paneru}. A separate normalization uncertainty is determined for each beam energy. These normalization uncertainties range from 4.1-9.8 $\%$.

\subsection{\texorpdfstring{$^3$H\MakeLowercase{e}$(\alpha,\gamma)^7$B\MakeLowercase{e}}{3He(a,g)} data}
\label{sec:capture_data}

The data selection~\cite{Kon13, Cos08, Bro07, Sin04, Car12, Bor13, Dil09} for the $^3$He$(\alpha,\gamma)^7$Be reaction for this work follows that of previous recent works~\cite{Ade11, deBoer:2014hha, Zhang:2019odg, Cyb08}. Note that the LUNA measurements of \citet{Gyu07} and \citet{Con07} are collected in \citet{Cos08}. The combined data sets cover a wide energy range from $E_{\text{c.m.}}$~=~94 to 3130~keV, but still remain below the proton decay threshold. Older data are not included due to a long history of discrepancies, which manifested as differences between experiments that used either direct detection of $\gamma$-rays or the activation technique. More recent measurements have achieved consistency resulting from improved experimental techniques by performing consistency check measurements using both direct detection of $\gamma$-rays and the activation technique~\cite{Ade11}.
Details about the capture data sets, including common-mode errors for cross sections, are listed in Table~\ref{tab:cmecapture}.

\subsection{Data Models} \label{sec:data_models}

Two distinct data models are analyzed here, $\mathcal{D}_{{\rm CS}}$ and $\mathcal{D}_{{\rm CSB}}$, where C indicates the inclusion of the capture data described in Sec.~\ref{sec:capture_data}, S indicates the inclusion of the SONIK data described in Sec.~\ref{sec:sonik_data}, and B indicates the inclusion of the Barnard data described in Sec.~\ref{sec:barnard_data}.
$\mathcal{D}_{{\rm CSB}}$ is a more complete data model in the sense that it includes more data and would naively be considered the ``best'' data model. But, there are notable effects when the data of \citet{Bar64} are included that are highlighted and discussed in Section~\ref{sec:results}.

\section{Results}
\label{sec:results}

The results of our analysis are presented here in two subsections.
The first discusses results in the energy regime of the data that was analyzed.
The second computes extrapolated quantities --- observables that lie in energy regimes outside those covered by the analyzed data.

\subsection{Fits to data}

First we examine the extent to which our results match experimental data.
We do this by comparing predicted and measured observables.

\subsubsection{Capture Data}

\begin{figure}
    \centering
    \includegraphics[width=\linewidth]{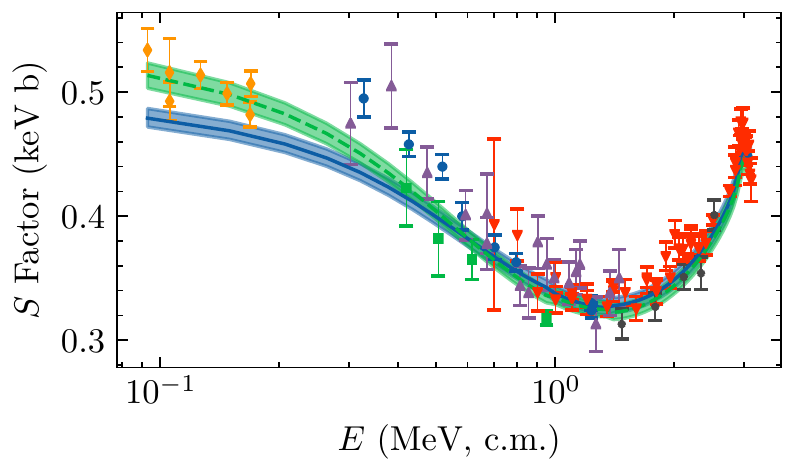}
    \caption{Total capture $S$ factor from Seattle~\cite{Bro18} (blue circles) Weizmann~\cite{Sin04} (green squares), LUNA~\cite{Cos08} (orange diamonds), ERNA~\cite{Dil09} (red, downward-pointing triangles), Notre Dame~\cite{Kon13} (purple, upward-pointing triangles), and ATOMKI~\cite{Bor13} (black stars) data sets are shown with reported error bars. $\mathcal{D}_{{\rm CSB}}$ and $\mathcal{D}_{{\rm CS}}$ results are shown with blue and green bands, respectively. The band indicates 68\% intervals. The solid, blue line indicates the median prediction from the $\mathcal{D}_{{\rm CSB}}$ analysis. The dashed, green line indicates the median prediction from the $\mathcal{D}_{{\rm CS}}$ analysis. Normalization factors have not been applied to either the theory prediction or data.}
    \label{fig:capture_fit_S}
\end{figure}

Figure~\ref{fig:capture_fit_S} shows the total capture $S$-factor data alongside bands representing 68\% intervals from the analyses of both data models, $\mathcal{D}_{{\rm CSB}}$ and $\mathcal{D}_{{\rm CS}}$.
For energies above 400~keV both analyses give very similar results. 
However, below that energy, the $\mathcal{D}_{{\rm CS}}$ analysis provides much better agreement with data. The LUNA data in particular discriminate between the two data models. The fit to the CSB data includes a normalization factor for the LUNA data that differs from 1 by about three times the stated common-mode error, cf. below. The normalization factors are not applied to the data in Fig.~\ref{fig:capture_fit_S}, which is why the CSB band sits well below the LUNA data.

\begin{figure}
    \centering
    \includegraphics[width=\linewidth]{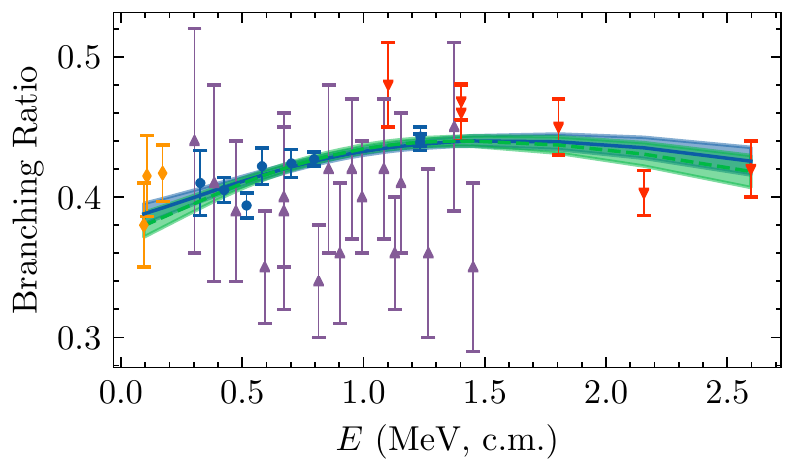}
    \caption{The branching ratio predictions are shown alongside the four analyzed branching ratio data sets: Seattle~\cite{Bro18}, LUNA~\cite{Cos08}, ERNA~\cite{Dil09}, and Notre Dame~\cite{Kon13}. Colors, symbols, and line styles are the same as Fig~\ref{fig:capture_fit_S}. Bands indicate 68\% intervals.}
    \label{fig:capture_fit_BR}
\end{figure}

Branching ratio results for both data models---$\mathcal{D}_{{\rm CS}}$ and $\mathcal{D}_{{\rm CSB}}$---are shown in Fig.~\ref{fig:capture_fit_BR}. The most prominent differences between the \DCSB and \DCS results occur near the upper and lower ends of the energy range. However, in the context of the experimental uncertainties, these differences are not significant.
Over the entire energy range, the predictions from \DCS and \DCSB overlap at the 1-$\sigma$ level.

\subsubsection{Scattering Data}

\begin{figure*}[t]
    \centering
    \includegraphics[width=\textwidth]{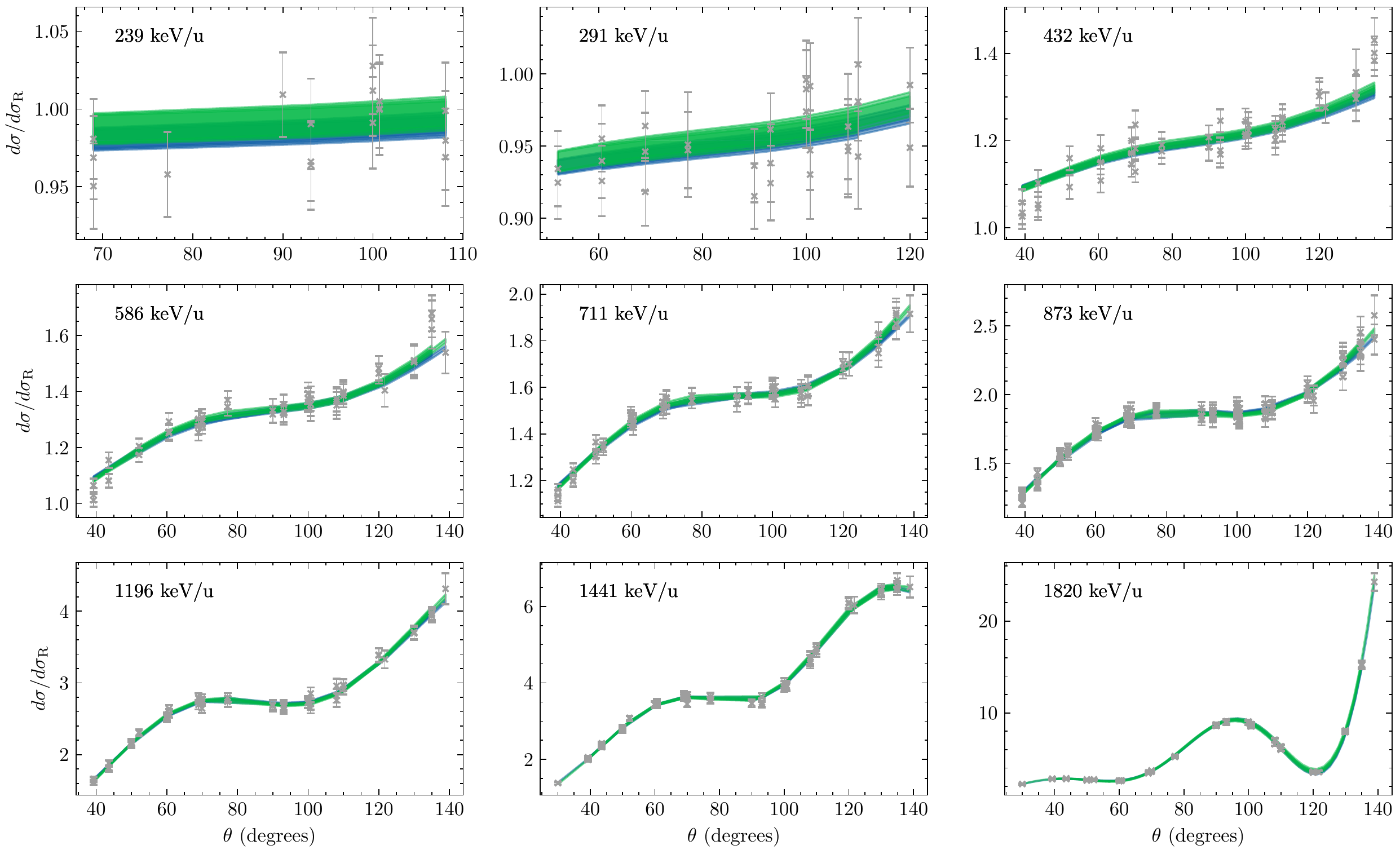}
    \caption{Angular dependence of the differential cross sections of \citet{the_Paneru} are shown relative to the Rutherford prediction with grey x's and error bars. Each panel includes the measurements from three interaction regions~\cite{PaneruThesis}. Bands indicate 68\% intervals. Green bands are generated for the analysis of \DCS. Blue bands correspond to \DCSB.}\label{fig:sonik_fit}
\end{figure*}

\begin{figure*}
    \centering
    \includegraphics[width=0.9\textwidth]{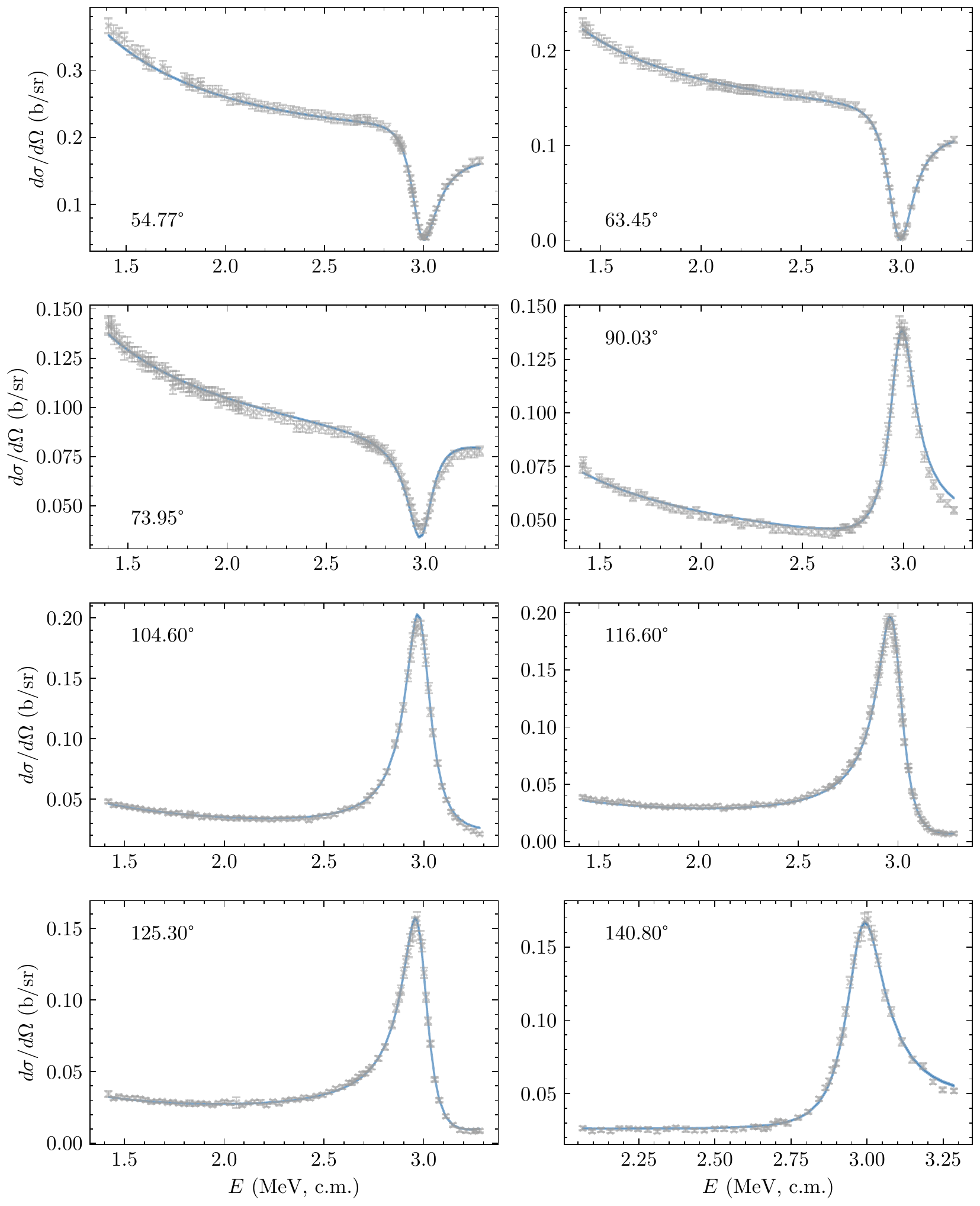}
    \caption{Differential cross section as a function of energy as reported in \citet{Bar64}, shown as grey x's with error bars. Blue bands represent the 68\% intervals generated from the \DCSB analysis.}
    \label{fig:barnard_fit}
\end{figure*}

The differential cross sections from the SONIK~\cite{the_Paneru} and \citet{Bar64} measurements are shown in Figs.~\ref{fig:sonik_fit} and \ref{fig:barnard_fit}, respectively, with the predictions from our analyses.
In all cases, both analyses reproduce the data to high accuracy.
However, the \DCS analysis results in a much lower $\chi^2/\text{datum}$ at $\max\ln P$: 0.72 for the SONIK~\cite{the_Paneru} data vs. 0.95 for the \DCSB analysis of the SONIK + Barnard~\cite{Bar64} data sets.


\subsection{Parameter Distributions}

Separate corner plots for each data model are provided in the Supplemental Material~\cite{supplemental_material}.
There are notable differences in several $R$-matrix parameters.
In particular, the \DCS ANCs are significantly larger and their posterior distributions are noticeably wider. The \DCS analysis also produces a significantly smaller ratio of ANCs, $C_1/C_0$. This is consistent with the smaller branching ratios at low energies shown in Fig.~\ref{fig:capture_fit_BR}.

The \DCS partial $\alpha$ widths in the 1/2${}^+$, 3/2${}^+$, and 5/2${}^+$ channels are smaller and separated by more than two standard deviations from the \DCSB widths.
The distributions for $\Gamma_{\gamma,0}^{(5/2^+)}$ seem to indicate opposite signs.
The \DCSB $E_x^{(7/2^-)}$ posterior is markedly smaller and narrower, and the constraints on $\Gamma_{\alpha}^{(7/2^-)}$ from \DCSB are dramatically tighter. This is presumably due to the much larger amount of data in the vicinity of the $7/2^-$ resonance that is present in the \citet{Bar64} data set.
It is also worth noting the ``non-Gaussian'' behavior of several of these distributions---a characteristic that would be difficult to identify in a typical analysis that assumed linear propagation of uncertainties around a minimum of the posterior pdf. Using Gaussian approximations and linearizing would likely underestimate uncertainties in the case of $\Gamma_{\gamma,0}^{(3/2^+)}$, for example.

\begin{figure*}[t]
    \centering
    \includegraphics[width=1.0\textwidth]{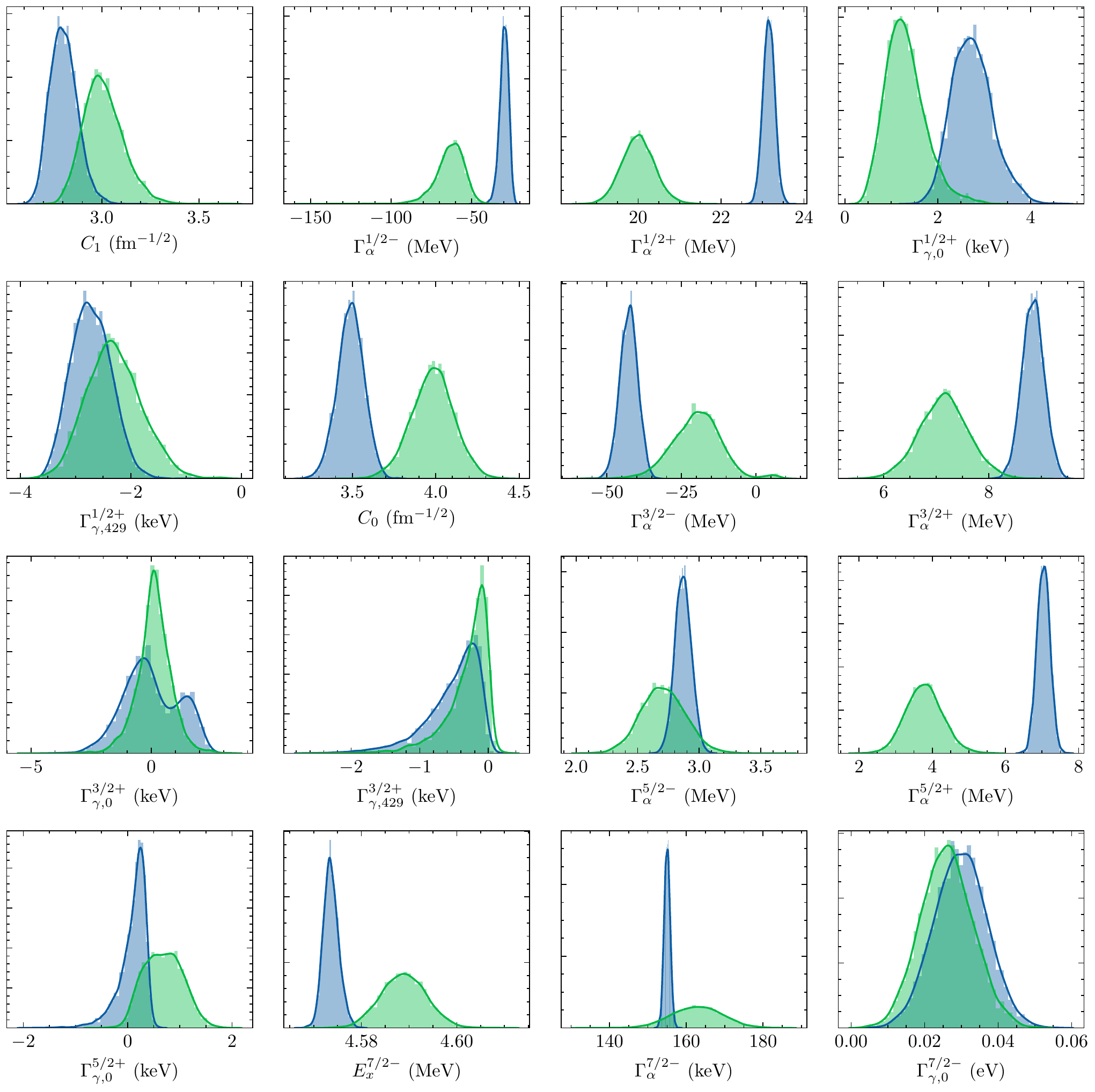}
    \caption{$R$-matrix parameter comparison between \DCS (green) and \DCSB (blue) analyses.}
    \label{fig:rpar_comparison}
\end{figure*}

All parameters shown in Fig.~\ref{fig:rpar_comparison} are well-constrained. By comparing to the prior distributions listed in Table~\ref{tab:sampled_rpars}, one can see the dominance of the data's influence over the information in the prior: all posterior distributions are markedly narrower than the priors chosen. As discussed in Sec.~\ref{sec:priors}, several $R$-matrix-model iterations were taken to remove redundant parameters.

\begin{figure*}[t]
    \centering
    \includegraphics[width=\textwidth]{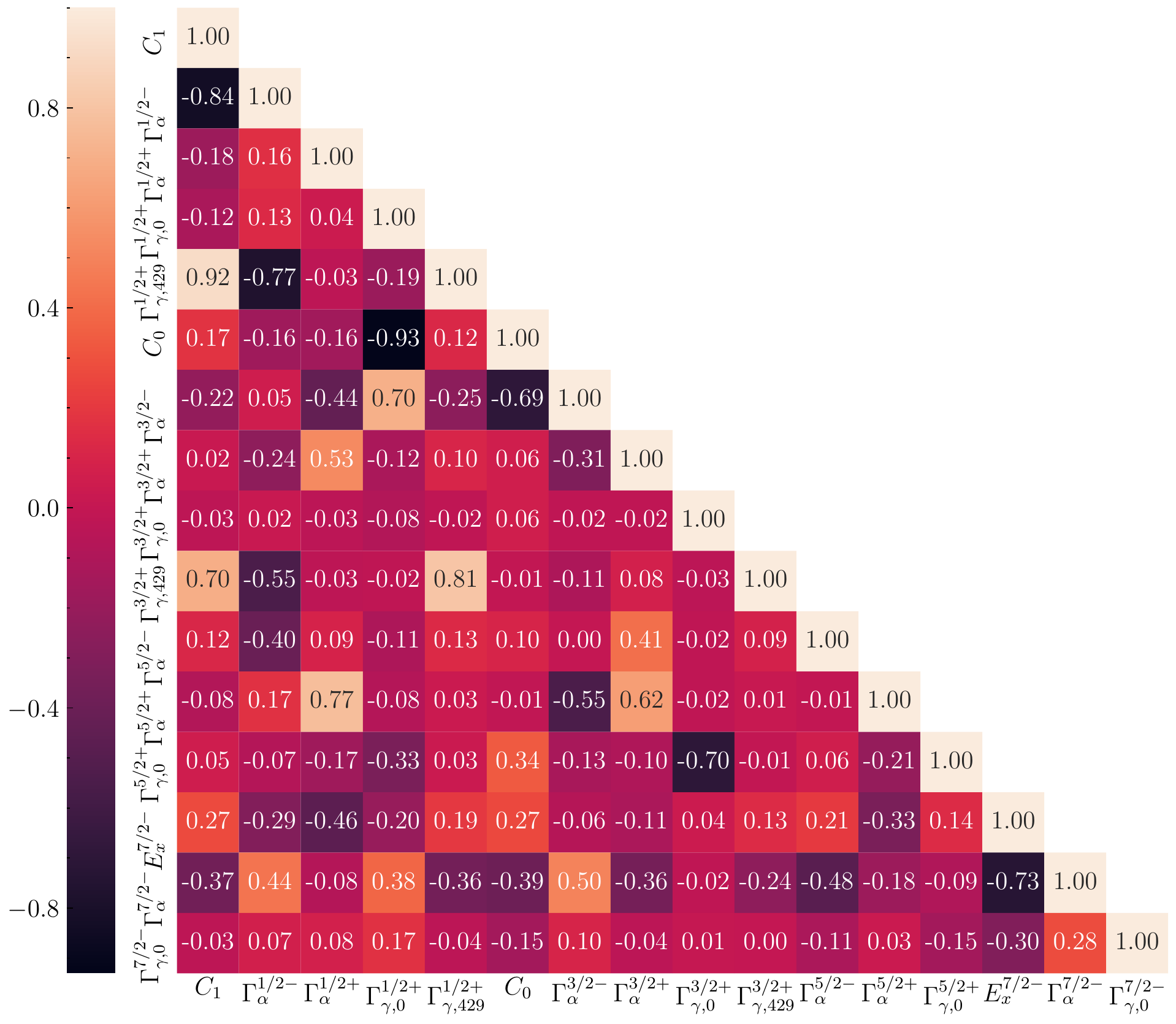}
    \caption{Correlation matrix of  $R$-matrix parameters for the \DCS analysis. Parameter chains are centered at zero and scaled to one prior to the computation. The strongest correlations (anti-correlations) are highlighted with lighter (darker) colors.}
    \label{fig:cov_matrix}
\end{figure*}

The correlation matrix of the $R$-matrix parameters is shown in Fig.~\ref{fig:cov_matrix}.
The figure represents an approximation of the full information contained in the corner plot given in the Supplemental Material~\cite{supplemental_material}. There, significant, often-nonlinear, correlations are observable between several pairs of $R$-matrix parameters.
In particular, the influence of the ANCs over the entire $R$-matrix parameter space, either directly or indirectly, 
means that it is very important for scattering data to have well-defined uncertainties over its full energy range. 

\begin{figure*}[t]
    \centering
    \includegraphics[width=\textwidth]{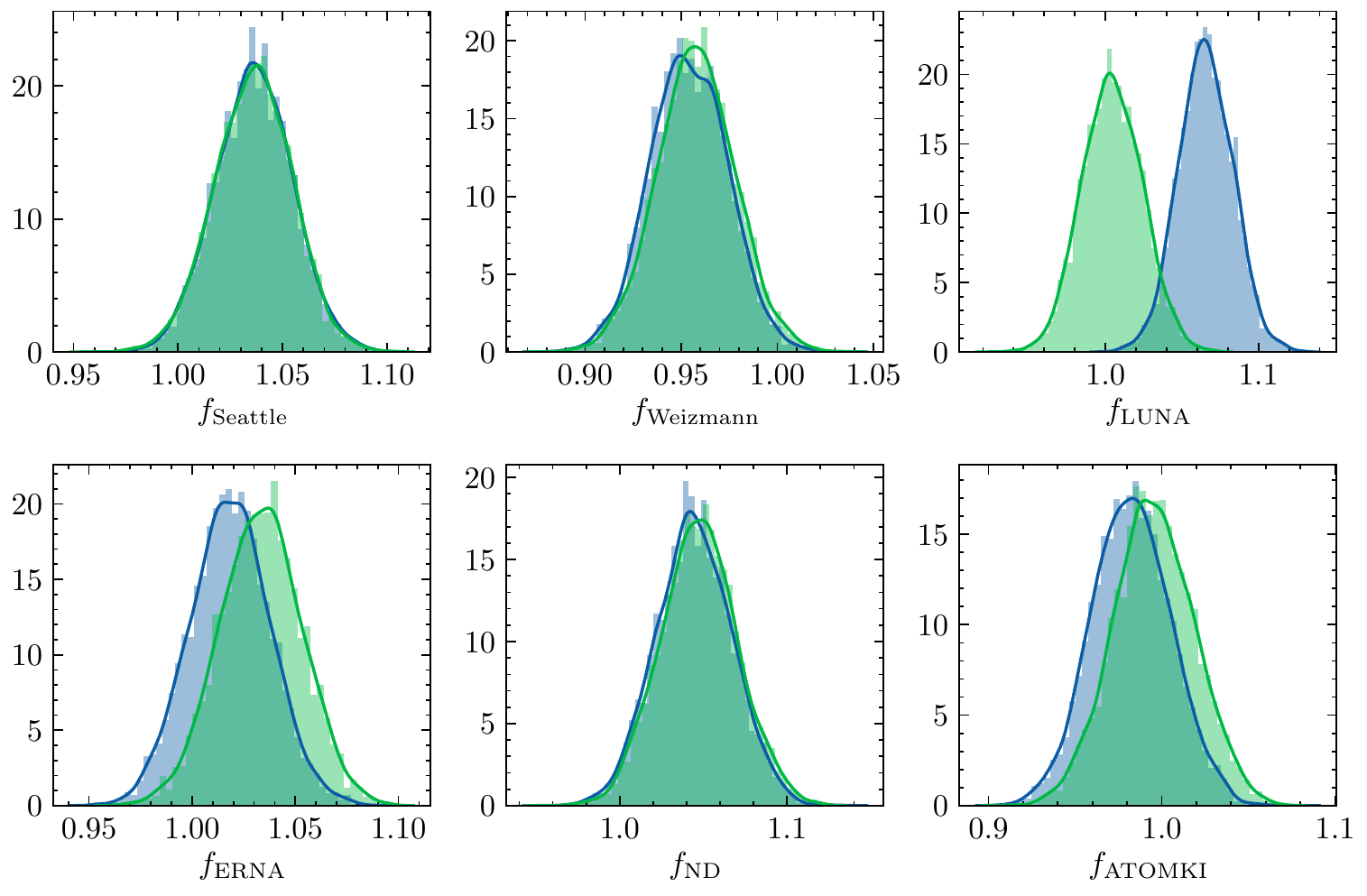}
    \caption{The normalization factors applied to the total cross section predicted by our $R$-matrix model are compared for each of the total capture data sets (Seattle~\cite{Bro18} Weizmann~\cite{Sin04}, LUNA~\cite{Cos08}, ERNA~\cite{Dil09}, Notre Dame~\cite{Kon13}, and ATOMKI~\cite{Bor13}). $\mathcal{D}_{{\rm CSB}}$ (blue) and $\mathcal{D}_{{\rm CS}}$ (green) results are shown together for each data set. Color online.}
    \label{fig:f_capture_comparison}
\end{figure*}

The normalization factors applied to the theory predictions for each of the total capture data sets are shown for both data models in Fig.~\ref{fig:f_capture_comparison}.
The comparison reveals good agreement between \DCS and \DCSB for all but the LUNA data set~\cite{Cos08}---the lowest-energy capture data set in our analysis.
The \DCS analysis yields a normalization factor for these data that is very close to 1.
In contrast, the \DCSB analysis requires that the LUNA data be shifted by nearly 10\%. (Recall from Eq.~(\ref{eq:likelihood}), that $f$ is applied to the theory prediction, and so an $f > 1$ corresponds to a systematic error that reduces the experimental cross section and uncertainties.) To put this in perspective, the LUNA collaboration estimates their common-mode error at 3.2\%. Because the LUNA data set is the lowest capture data set, this disagreement between the \DCS and \DCSB analyses corresponds to a significant difference in the extrapolated $S(0)$ of these two analyses.

\begin{figure}
    \centering
    \includegraphics[width=\linewidth]{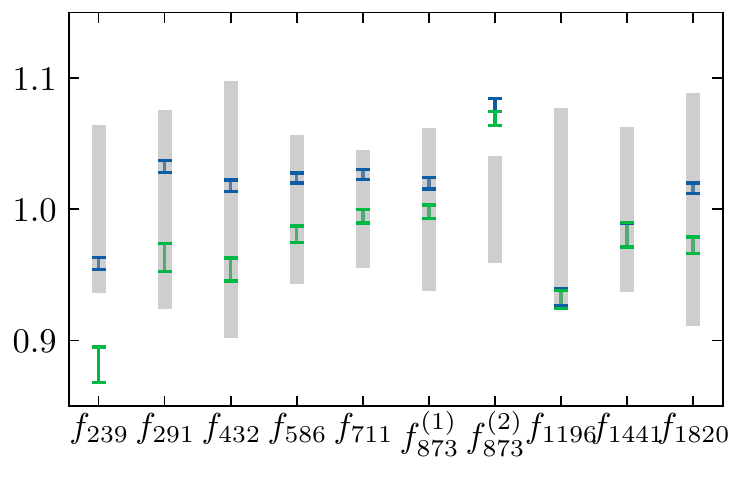}
    \caption{Summaries of the normalization factor posteriors for each SONIK~\cite{the_Paneru} data set are shown for $\mathcal{D}_{{\rm CSB}}$ (blue) and $\mathcal{D}_{{\rm CS}}$ (green). Error bars represent 68\% quantiles. Grey-shaded rectangles indicate the uncertainties reported in~\cite{PaneruThesis}.}
    \label{fig:f_sonik_comparison}
\end{figure}

The normalization factors applied to the theory predictions for each of the SONIK energies are shown in Fig.~\ref{fig:f_sonik_comparison}. When the data of \citet{Bar64} are included in the analysis, the SONIK normalization factors are significantly larger. This effect is systematically apparent at lower energies. In more than half the cases, the \DCSB and \DCS results are inconsistent with each other. For eight out of ten SONIK energies, the normalization factor obtained from the fit is within the common-mode error estimated by the SONIK collaboration. Note that the common-mode error in this experiment  was estimated to be different at different beam energies~\cite{PaneruThesis}~\footnote{We use slightly different common-mode uncertainty estimates in our prior definitions than those listed in Ref.~\cite{PaneruThesis}.
This update will be reflected in a forthcoming publication by the SONIK collaboration~\cite{the_Paneru}}.
This is represented in Fig.~\ref{fig:f_sonik_comparison} by the varying heights of the grey bands, which are priors in accord with these experimentally assigned common-mode errors, see Table~\ref{tab:cmesonik}.

\begin{figure*}[t]
    \centering
    \includegraphics[width=\textwidth]{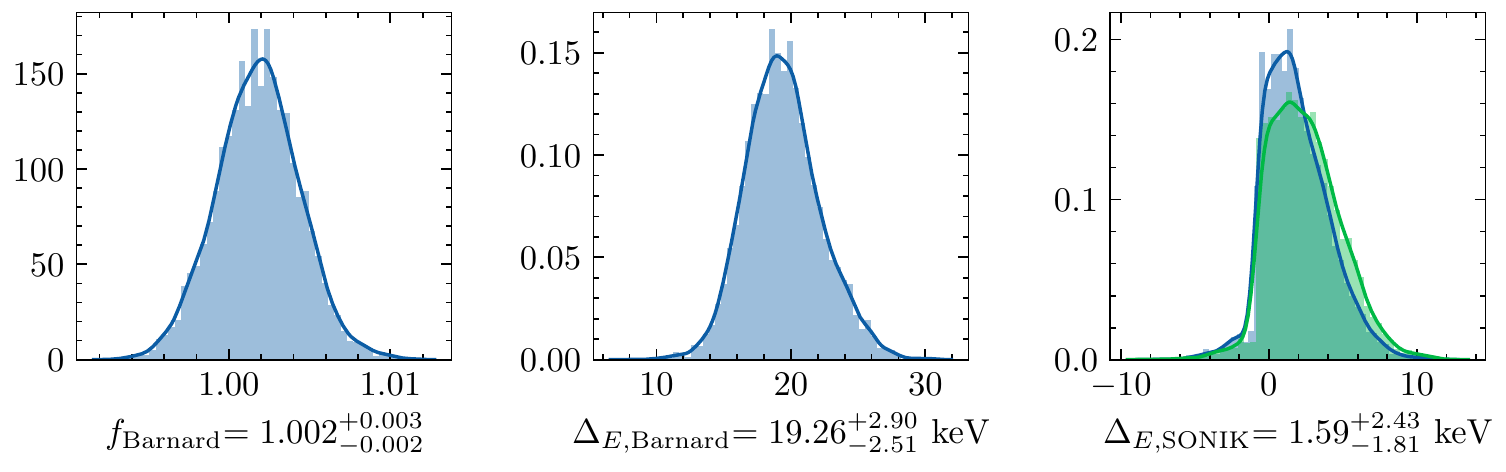}
    \caption{Posteriors of the normalization factor applied to the  Barnard data and the energy shifts introduced to the \citet{Bar64} and SONIK~\cite{the_Paneru} data sets. The Barnard normalization factor is applied to the theory prediction. Energy shifts are presented in keV. These results were obtained exclusively with the $\mathcal{D}_{{\rm CSB}}$ data model.}
    \label{fig:f_Barnard_energy_shifts}
\end{figure*}

The posteriors for $f_{\rm Barnard}$ and the energy shifts for both the \citet{Bar64} and SONIK~\cite{the_Paneru} data sets (see Sec.~\ref{sec:systematic_errors}) are shown in Fig.~\ref{fig:f_Barnard_energy_shifts}.
The result for $f_{\rm Barnard}$ is $1.002^{+0.003}_{-0.002}$: well within the estimated systematic uncertainty of 5\% given in \citet{Bar64}.
A shift of $19.26^{+2.90}_{-2.51}$~keV in the energies reported in \citet{Bar64} is found, but this result is consistent with the energy uncertainty estimates ranging from 20-40~keV given in that paper. However, even such a clearly nonzero shift does not seem to significantly impact extrapolated quantities. 
Finally, the SONIK energy shift indicated by our analyses is $1.59^{+2.43}_{-1.81}$~keV.
This result matches very well with the reported energy uncertainty estimate of 3~keV.
The prior for this parameter was a normal distribution centered at 0~keV with a 1-$\sigma$ width of 3~keV.
The primary difference between the posterior and the prior for this parameter is the loss of probability in the negative energy region.
If any energy shift in the SONIK data~\cite{the_Paneru} is necessary, it is positive, but since 0~keV is well within one standard deviation, there is strong evidence for no shift.

The ANCs corresponding to the two bound ${}^7{\rm Be}$ states are of particular interest for extrapolating threshold quantities.
First, we point out that the inclusion of scattering data significantly reduces the uncertainty of the ANCs. Our posterior is much narrower than that obtained using capture-only data in Ref.~\textcite{Zhang:2019odg}. 
This highlights the importance of scattering data in constraining bound-state properties and the amplitudes associated with transitions to them.

\begin{figure}
    \centering
    \includegraphics[width=\linewidth]{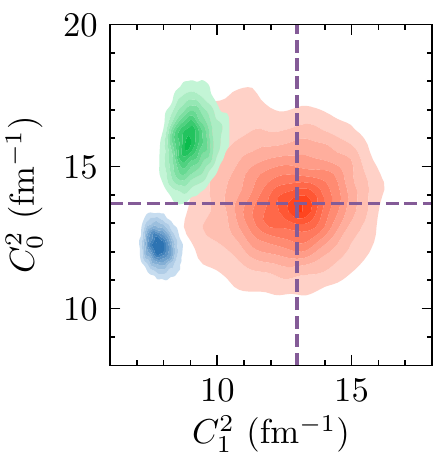}
    \caption{The two-dimensional posterior of the squares of the ANCs, $C_0$ and $C_1$. Results for $\mathcal{D}_{{\rm CS}}$ are shown in green and for $\mathcal{D}_{{\rm CSB}}$ in blue. The EFT analysis of capture data of \citet{Zhang:2019odg} extracted the ANC values shown in red, and in the analysis of \citet{Bar64} and capture data of \citet{deBoer:2014hha} the ANCs were fixed at the location indicated by the purple, dashed lines.}
    \label{fig:anc_comparison}
\end{figure}

Second, the choice of scattering data set matters. The $C_1$ results from analyzing \DCS\ and \DCSB\ are discrepant at the 1-$\sigma$ level.
The $C_0$ results disagree by approximately 2-$\sigma$.
The contrast is highlighted in Fig.~\ref{fig:anc_comparison} where the squares $C_1^2$ and $C_0^2$ are compared. The differing values directly impact the $S$-factor extrapolations discussed below.

\subsection{Extrapolated quantities}


The Coulomb-modified effective range function is given in~\citet{HAMILTON1973443} and~\citet{Haeringen1977} as
\begin{equation}
    \label{eq:coulomb_ere}
    K(E) = k^{2\ell+1} \frac{\eta^{2\ell}}{\Gamma^2(\ell+1)}u_\ell(\eta)\left[C_0^2(\eta)\cot\delta_\ell+2\eta h(\eta)\right]~,
\end{equation}
where $k$ is the relative momentum, $\ell$ is the angular momentum, $\eta$ is the Sommerfeld parameter, $\Gamma$ is the gamma function, $u_\ell(\eta)$ is given by
\begin{equation}
       u_\ell(\eta) = \frac{ \Gamma^2(2\ell+2) C_\ell^2 }{ (2\eta)^{2\ell}C_0^2 }~,\\
\end{equation}
with
\begin{eqnarray}
    C_0 &=& \left[\frac{2\pi\eta}{e^{2\pi\eta}-1}\right]^{1/2}~,\\
    h(\eta) &=& \frac{1}{2} \left[\Psi(1+i\eta) + \Psi(1-i\eta)\right] - \ln\eta~,
\end{eqnarray}
and $\Psi$ representing the digamma function~\cite{Humblet1985}.
This effective range function is an analytic function of $E$ (or $k^2$) near $E=0$. From the phase shifts, obtained with \texttt{BRICK}, calculated over a range of low momenta, one can fit the scattering length, $a_0$, and effective range, $r_0$, according to the low-energy expansion
\begin{equation}
    \label{eq:ere_expansion}
    K(E) = -\frac{1}{a_0} + \frac{r_0}{2} k^2 + \dots
\end{equation}
Our calculation involves 70 equally spaced phase shifts over a range of energies from 0.57 keV to 3.93 MeV.
The results are used to evaluate the effective range function defined by \eqref{eq:coulomb_ere}.
The energy dependence is then fit to \eqref{eq:ere_expansion} using a non-linear least squares fit.

\begin{figure}
    \centering
    \includegraphics[width=\linewidth]{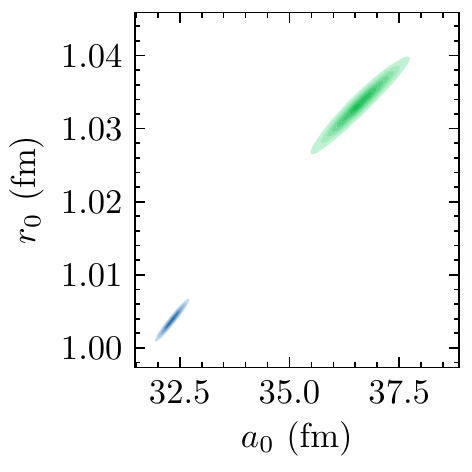}
    \caption{$a_0$-$r_0$ correlation for both $\mathcal{D}_{{\rm CSB}}$ (blue) and $\mathcal{D}_{{\rm CS}}$ (green) data models. Color online.}
    \label{fig:a0_r0_comparison}
\end{figure}

The results from $\mathcal{D}_{{\rm CSB}}$ and $\mathcal{D}_{{\rm CS}}$ are shown in Fig.~\ref{fig:a0_r0_comparison}.
As in the ANC comparison, they are strikingly discrepant. The naive expectation would be that \DCSB distributions would be smaller subsets of the \DCS distributions.
For many relevant quantities, this is not the case.

\begin{figure}
    \centering
    \includegraphics[width=\linewidth]{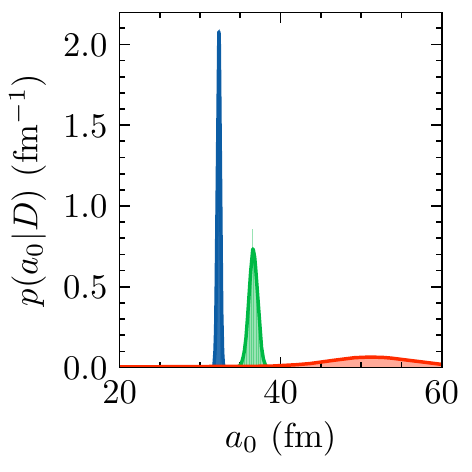}
    \caption{$a_0$ posteriors obtained from \DCS (green) and \DCSB (blue) analyses. The result from \citet{Zhang:2019odg} is shown in red.}
    \label{fig:a0_comparison}
\end{figure}

Figure~\ref{fig:a0_comparison} shows a comparison of the scattering lengths obtained from the \DCS and \DCSB analyses.
A comparison to \citet{Zhang:2019odg}, also included in Figure~\ref{fig:a0_comparison}, reveals the impact of including scattering data: the inclusion of scattering data drives the median downward and constrains the uncertainties significantly.
A summary of these posteriors is given in Table~\ref{tab:extrapolated_quantities}.

The \DCSB scattering length and effective range are both smaller and more tightly constrained. One might have expected that with more data---and more data at lower energies---this extrapolated quantity would become more tightly constrained.
The two-dimensional posteriors shown in Fig.~\ref{fig:a0_r0_comparison} seem to lie on the same line or band that defines the correlation between $a_0$ and $r_0$, though two extended posteriors is not sufficient to define such a line.

\begin{figure}
    \centering
    \includegraphics[width=\linewidth]{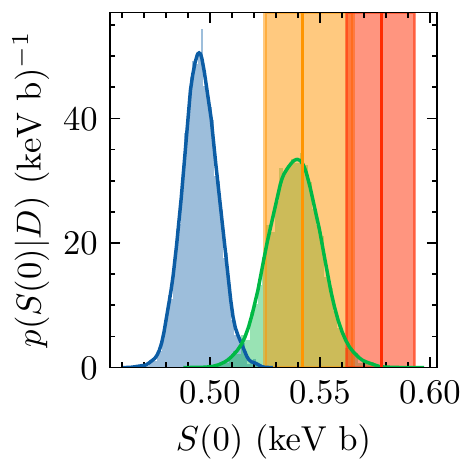}
    \caption{Extrapolated $S(0)$ posteriors from the analyses of both $\mathcal{D}_{{\rm CSB}}$ (blue) and $\mathcal{D}_{{\rm CS}}$ (green) data models. Previous results from \citet{Zhang:2019odg} (red) are \citet{deBoer:2014hha} (orange) are also summarized here for comparison.}
    \label{fig:s0_comparison}
\end{figure}

The total capture $S$ factor at zero energy was extrapolated by evaluating the $S$ factor at 100 evenly spaced points between 1 to 100~keV, constructing a cubic-polynomial interpolation function to represent the calculations, and evaluating that function at zero energy.
Errors from the interpolation/extrapolation process are negligible when compared to contributions from parameter uncertainties. The results are shown alongside previous results in Fig.~\ref{fig:s0_comparison}. As expected from the different low-energy behaviors shown in Fig.~\ref{fig:capture_fit_S}, the $\mathcal{D}_{\rm CS}$ and $\mathcal{D}_{\rm CSB}$ results are discrepant, only overlapping at the 2-$\sigma$ level. The inclusion of the \citet{Bar64} data reduces the uncertainty in $S(0)$ and pulls the entire distribution downward, outside the uncertainties of the \DCS analysis.
This effect is not seen in~\cite{deBoer:2014hha} because the ANCs in that analysis were not varied freely. The \DCSB result is discrepant with the \DCS results \textit{and} those reported in~\cite{deBoer:2014hha} and~\cite{Zhang:2019odg}.
A summary of these posteriors is given in Table~\ref{tab:extrapolated_quantities}.

\begin{table*}[]
    \centering
    \begin{ruledtabular}
    \begin{tabular}{c|c|c|c}
         Analysis & $S(0)$ (keV b) & $a_0$ (fm) & $r_0$ (fm) \\
         \hline
         \DCS & $0.539^{+0.011}_{-0.012}$ & $36.59^{+0.55}_{-0.53}$ & $1.033^{+0.003}_{-0.003}$ \\
         \DCSB & $0.495^{+0.008}_{-0.008}$ & $32.32^{+0.18}_{-0.18}$ & $1.004^{+0.001}_{-0.001}$ \\
         \citet{deBoer:2014hha} & $0.542^{+0.023}_{-0.017}$ & --- & --- \\
         \citet{Zhang:2019odg} & $0.578^{+0.015}_{-0.016}$ & $50.36^{+6.02}_{-7.50}$ & $0.974^{+0.025}_{-0.027}$
    \end{tabular}
    \caption{A summary of the posteriors of the extrapolated quantities. Where possible, results from other anlayses are included.}
    \label{tab:extrapolated_quantities}
    \end{ruledtabular}
\end{table*}

\begin{figure}
    \centering
    \includegraphics[width=\linewidth]{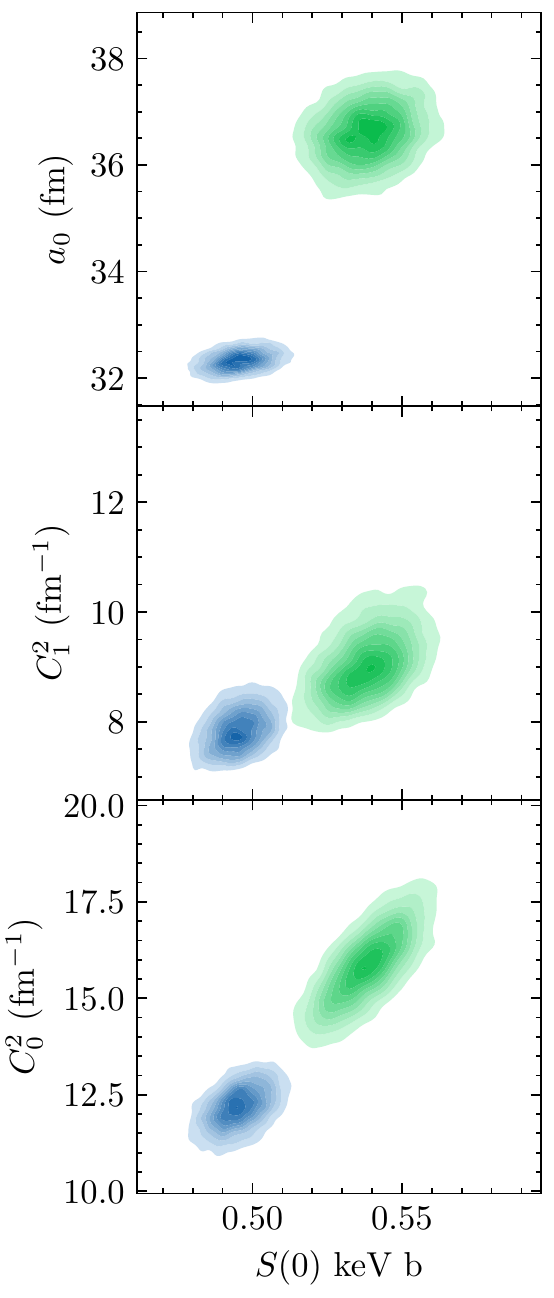}
    \caption{Two-dimensional posteriors are presented for the analyses of both $\mathcal{D}_{{\rm CSB}}$ (blue) and $\mathcal{D}_{{\rm CS}}$ (green) data models. The ``anchor'' parameter is $S(0)$. The top panel gives its correlation with $a_0$. The middle (bottom) panel corresponds to the square of the excited- (ground-) state ANC.}
    \label{fig:s0_a0_c1sq_c0sq_correlation}
\end{figure}

Insights into the relevance of parameters can be obtained by examining the correlations between them. In Fig.~\ref{fig:s0_a0_c1sq_c0sq_correlation}, the correlations between $S(0)$ and $a_0$, $C_1^2$ and $C_0^2$ are shown. While the \DCS and \DCSB results are discrepant in several astrophysically relevant cases, the discrepancy is consistent, and this figure exposes, to a large extent, why: the ANCs, particularly the ground-state ANC, strongly correlates with $S(0)$. The \citet{Bar64} data more tightly constrain these parameters at smaller values, and this directly lowers the predicted $S(0)$ extrapolation.

\section{Conclusions}

We have described and applied the Bayesian $R$-matrix Inference Code Kit (\texttt{BRICK}), which facilitates communication between the phenomenological $R$-matrix code \texttt{AZURE2}~\cite{Azu10} and a Markov Chain Monte Carlo (MCMC) sampler such as \texttt{emcee}~\cite{For13}. It thereby enables MCMC sampling of the joint posterior probability density function (pdf) for the $R$-matrix parameters and normalization factors. With samples that represent such a posterior in hand, the computation of the pdf for any  quantity that can be calculated in the $R$-matrix formalism is straightforward.
    
While \texttt{BRICK} is a general tool, we have also provided an example of its application to an $R$-matrix fit of ${}^3$He-$\alpha$ scattering and the ${}^3$He$(\alpha,\gamma)^7$Be capture reaction data, in order to make inferences about the ${}^7$Be system. This application was partly motivated by the availability of a new ${}^3$He-$\alpha$ scattering data set obtained using the SONIK detector at TRIUMF~\cite{PaneruThesis} following the suggestion of \citet{deBoer:2014hha}. These data have more carefully quantified uncertainties than a previous measurement by \citet{Bar64}. Our study shows this motivation was well justified, finding discrepant values for extrapolated quantifies when the data of \citet{Bar64} were included. 
Our analysis of the SONIK data shows consistency between them and capture data, producing an $S$ factor in accord with analyses of capture data alone: our final \DCS (capture + SONIK data) result for the $S$-factor at zero energy is $S(0)=0.539^{+0.011}_{-0.012}~{\rm keV}~{\rm b}$. When the \citet{Bar64} data were included in the analysis, the \DCSB results produced significantly lower ANCs and $S(0)$ extrapolation. Indeed, the \DCSB analysis produces values for $S(E)$ at c.m. energies of 10--20~keV that can only be reconciled with the LUNA data~\cite{Cos08} if the normalization of these data is adjusted by 2--3 times the quoted common-mode error. 
    
This emphasizes the importance of detailed uncertainty quantification when data sets are to be used for accurate inference of extrapolated quantities, where \citet{Bar64} does not include these kinds of details regarding the experiment. This makes the tension between the \citet{Bar64} and SONIK data regarding $S(0)$ difficult to resolve, thus the \citet{Bar64} data may need to be omitted from future evaluations. We emphasize, though, that these previous data were invaluable in advancing our understanding of the ${}^7$Be system to its current state, but data with more well defined uncertainties are needed for current applications.     

Zhang, Nollett, and Phillips pointed out that the $s$-wave ${}^3$He-$\alpha$ scattering length is correlated with this result~\cite{Zhang:2019odg}. The \DCS analysis produces $a_0=36.59^{+0.55}_{-0.53}$~fm.
Premarathna and Rupak simultaneously analysed capture data and  ${}^3$He-$\alpha$ phase shifts in EFT and found $a_0=40^{+5}_{-6}$ fm (Model A II of Ref.~\cite{Premarathna:2019tup})---in good agreement with this number.
However, it disagrees by 2$\sigma$ with the $a_0$ extracted using EFT methods from capture data alone by \citet{Zhang:2019odg}: $a_0=50^{+6}_{-7}$. Recently ~\citet{Poudel:2021mii} performed an EFT analysis of the SONIK data, using priors on the ${}^7$Be ANCs from the capture analyis of Ref.~\cite{Zhang:2019odg}, and extracted $a_0=60 \pm 6$ fm---even further away from the results of this $R$-matrix analysis.

Improvements in the analyses presented here could occur if there were:
\begin{itemize}
    \item Better documentation of the energy dependence of systematic uncertainties in published data sets. The Bayesian formalism that underlies \texttt{BRICK} allows systematic uncertainties with any correlation structure to be incorporated into the analysis. 
    
    \item Improved understanding of the way theory uncertainties in the phenomenological $R$-matrix formalism affect the extrapolation of data.
    
    \item Detailed modern data with full uncertainty quantification in the vicinity of the $7/2^-$ resonance. This may help resolve some of the ambiguities in results between the \DCS and \DCSB analyses.
    
    \item Ab initio constraints, e.g., on ANCs could be incorporated in the analysis.
    
    \item Data from transfer reactions that provided complementary information on the ${}^7$Be ANCs.
        \end{itemize}
        
Future applications of \texttt{BRICK} could include posteriors for astrophysical reaction rates.
This would enhance \texttt{BRICK}'s utility as a tool for performing detailed uncertainty quantification on nuclear reactions, especially those of astrophysical interest.
\texttt{AZURE2} already includes the necessary functionality. Implementing this feature ought to be a straightforward process.

\begin{acknowledgments}

We thank Maheshwor Poudel and Xilin Zhang for helpful discussions. This work was supported by the U.S. Department of Energy, National Nuclear Security Agency under Award DE-NA0003883,
and Office  of  Science,  Office  of  Nuclear  Physics, under  Award DE-FG02-93ER-40756, as well as by
the National Science Foundation under grants OAC-2004601 (CSSI program, BAND collaboration),
PHY-2011890 (University of Notre Dame Nuclear Science Laboratory), and PHY-1430152 (the Joint Institute for Nuclear Astrophysics - Center for the Evolution of the Elements). This research utilized resources from the Notre Dame Center for Research Computing.  
\end{acknowledgments}

\section{Supplemental Material}

\begin{figure*}
    \centering
    \includegraphics[width=1.0\textwidth]{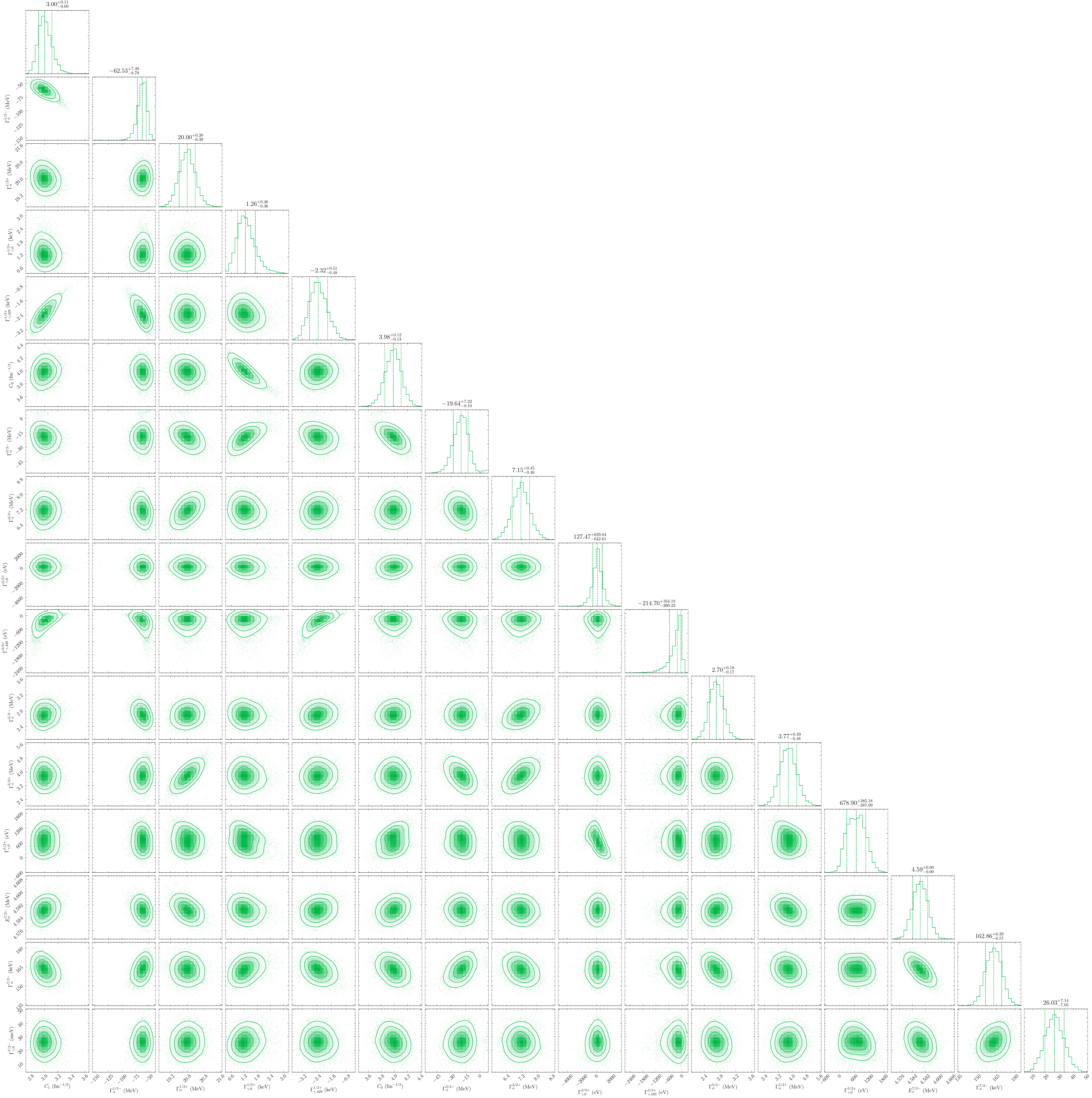}
    \caption{Posterior distributions of the $R$-matrix parameters sampled in the \DCS analysis.}
    \label{fig:rpar_corner_cs}
\end{figure*}

\begin{figure*}
    \centering
    \includegraphics[width=1.0\textwidth]{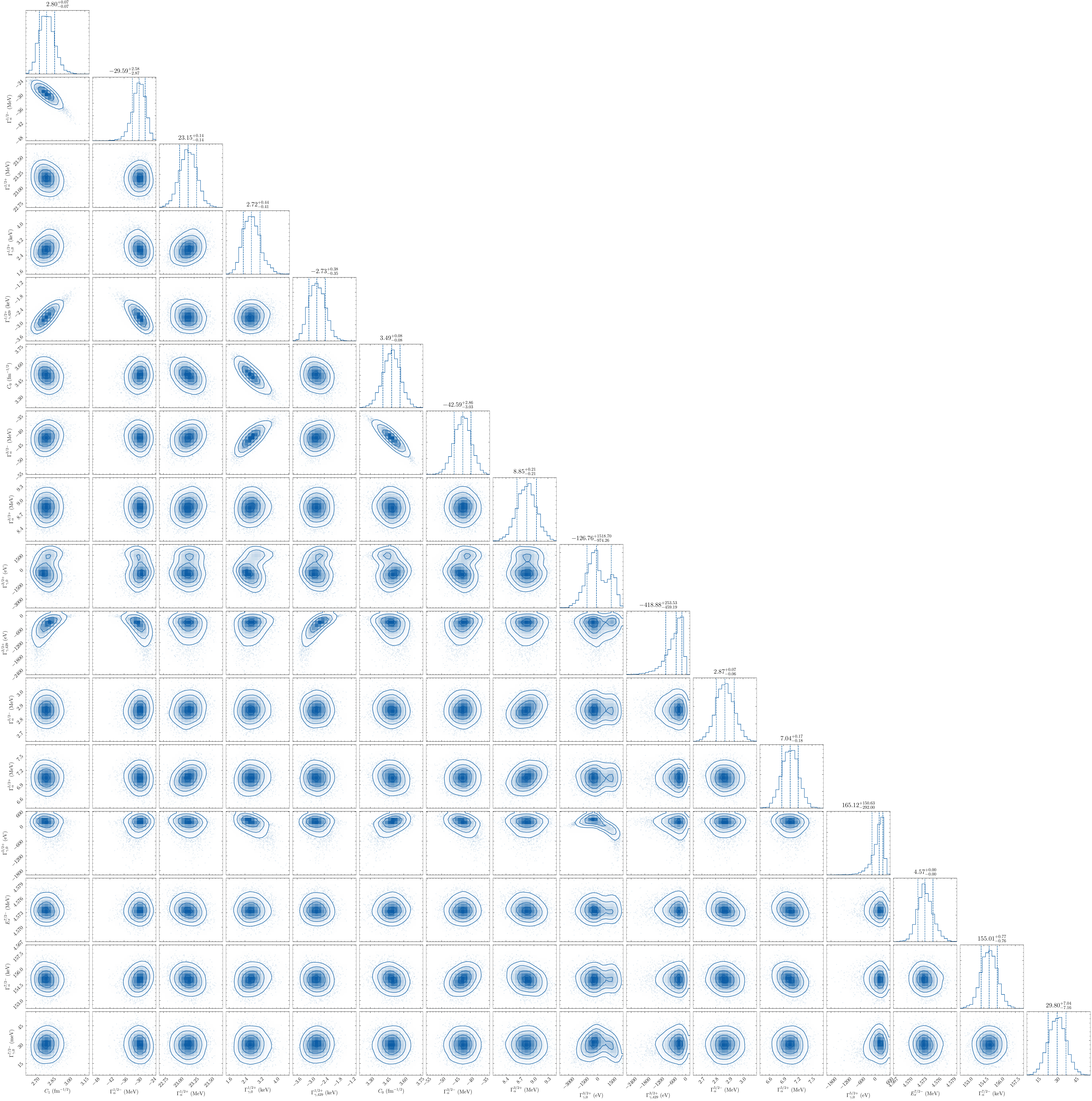}
    \caption{Posterior distributions of the $R$-matrix parameters sampled in the \DCSB analysis.}
    \label{fig:rpar_corner_csb}
\end{figure*}

Posterior probability distributions resulting from the \DCS and \DCSB analyses are shown in Figures~\ref{fig:rpar_corner_cs} and~\ref{fig:rpar_corner_csb}, respectively.

\label{sec:supplemental_material}

\end{document}